\documentclass{pspum-l}

\usepackage{amssymb}

\usepackage{graphicx}

\usepackage{physics}
\usepackage{tikz} 
\usepackage{amsmath}

\makeatletter
\newcommand{\smallbullet}{} 
\DeclareRobustCommand\smallbullet{%
  \mathord{\mathpalette\smallbullet@{0.5}}%
}
\newcommand{\smallbullet@}[2]{%
  \vcenter{\hbox{\scalebox{#2}{$\m@th#1\bullet$}}}%
}
\makeatother

\newtheorem{theorem}{Theorem}[section]

\theoremstyle{definition}

\theoremstyle{remark}

\numberwithin{equation}{section}

\begin{document}

\title{Giant Gravitons and non-conformal vacua \\ in twisted holography}


\author{Kasia Budzik}
\address{Perimeter Institute for Theoretical Physics, Waterloo, ON N2L 2Y5, Canada}
\curraddr{}
\email{kbudzik@perimeterinstitute.ca}
\thanks{}


\date{}

\begin{abstract}
Twisted holography relates the two-dimensional chiral algebra subsector of $\mathcal{N}=4$ SYM to the B-model topological string theory on the deformed conifold $SL(2,\mathbb{C})$. We review the relevant aspects of the duality and its two generalizations: the correspondence between determinant operators and ``Giant Graviton" branes and the extension to non-conformal vacua of the chiral algebra.
\end{abstract}

\maketitle


\section{Introduction and summary}

\subsection{Twisted holography}

Twisting supersymmetric quantum field theories is the procedure of passing to the cohomology of a fermionic supercharge $\mathcal{Q}$. The operation produces a consistent subsector of the SQFT which is easier to study and may allow for exact computations. In the twisted theory several simplifications occur:
\begin{itemize}
    \item Twisting restricts to protected (BPS) operators. Certain quantities, eg. correlation functions, might become independent of various parameters like the coupling constant.
    
    \item Dependence on certain spacetime coordinates drops out in cohomology. The fermionic supercharges satisfy anticommutation relations of the sche\-matic form
    \begin{align}
        \{ Q,\tilde Q \} \sim P \, .
    \end{align}
    The momenta that appear in the image of the twisting supercharge $\{\mathcal{Q},\,\smallbullet\,\} \sim P$, are by definition exact in cohomology. Since momenta generate spacetime translations, dependence on the corresponding spacetime coordinates drops out in the twisted theory. There might be many twists available depending on the spacetime dimension and the number of supersymmetry (see eg. \cite{Elliott:2018cbx,Elliott:2020ecf}). For example, if all of the spacetime momenta become exact we have \emph{the topological twist}, and if half -- \emph{the holomorphic twist}.

    \item Twisted theories are endowed with extra mathematical structures and symmetries, which become visible at the level of cohomology \cite{Budzik:2022mpd,4dcohomology,Garner:2022its,Saberi:2019fkq,Saberi:2019ghy,Gwilliam:2018lpo}. However, this topic is beyond the scope of this note.
\end{itemize}

For the supersymmetric theories that have a holographic dual, a natural question is whether there exists an analogous twisting procedure on the gravity side. It was proposed in \cite{Costello:2016mgj} that the dual operation corresponds to turning on a non-zero value for a bosonic ghost field\footnote{The role of the bosonic ghost field in the BV formalism is to gauge a local supersymmetry. Coupling a supersymmetric field theory (worldvolume theory of a brane) to the background where a bosonic ghost $\Psi$ is non-zero is equivalent to adding the corresponding supercharge $Q_\Psi$ to the BRST charge of the supersymmetric field theory.} of the supergravity theory. The twisted type IIB supergravity was conjectured to be equivalent to the BCOV theory.\footnote{The stronger conjecture is that twisted type IIB superstring theory is equivalent to topological B-model.} Remarkably, twisted supergravity can be quantized to all orders in perturbation theory, despite the theory being non-renormalizable \cite{Costello:2015xsa,Costello:2016nkh}.
    
Applying the twisting procedures to both sides of a holographic duality should therefore produce an easier and more tractable duality, where many simplifications occur.\footnote{There are also examples (holomorphic twist of $\mathcal{N}=1$ SYM and SQCD) where the dual of the original theory is not known, yet one can conjecture the holographic dual of the twisted theory \cite{4dcohomology}.} In addition, objects that appear after twisting are typically more well-defined mathematically (eg. vertex algebras, BCOV theory). One could hence hope for a more mathematical formulation of holography, at least at the twisted level. It has been proposed that twisted holography is captured by the ubiquitous mathematical duality known as Koszul duality \cite{Costello:2017fbo, Costello:2020jbh, Paquette:2021cij, Zeng:2023qqp, Costello:2022wso}.

The main example\footnote{Other examples of twisted supergravity/holography include \cite{Eager:2021ufo,Raghavendran:2021qbh,Brunner:2021tfl,Saberi:2021weg,Raghavendran:2019zdq,Ishtiaque:2018str,Gaiotto:2020vqj,Oh:2020hph,Oh:2021wes,Oh:2021bwi,Gaiotto:2020dsq,Costello:2021kiv,Li:2019qzx,Costello:2022wso,Costello:2022jpg,Zhou:2022mzb}.} of twisted holography in this note is the duality between a 2d chiral algebra $\mathcal{A}_N$ and the topological B-model on $SL(2,\mathbb{C})$ proposed in \cite{Costello:2018zrm}. The chiral algebra is a subsector of $\mathcal{N}=4$ SYM obtained by the ``$Q+S$" twist of \cite{Beem:2013sza}.\footnote{For an earlier work on the holographic dual of the chiral algebra subsector of $\mathcal{N}=4$ SYM see \cite{Bonetti:2016nma}.} One can arrive at the dual geometry $SL(2,\mathbb{C})$ by considering a stack of D1-branes in the topological B-model on $\mathbb{C}^3$, whose worldvolume theory is precisely the chiral algebra $\mathcal{A}_N$ (in analogy to the original derivation by Maldacena \cite{Maldacena:1997re}). The stack of branes deforms the complex structure $\mathbb{C}^3\setminus \mathbb{C}$ to $SL(2,\mathbb{C})$. According to \cite{Costello:2018zrm}, B-model on $SL(2,\mathbb{C})$ can also be seen as a twist of type IIB string theory on $\textrm{AdS}_5\times S^5$, where $SL(2,\mathbb{C})\cong \textrm{AdS}_3\times S^3 \subset \textrm{AdS}_5\times S^5$. Therefore, the above duality can be regarded as a subsector of the standard $\textrm{AdS}_5/\textrm{CFT}_4$ correspondence.

At the twisted level, many aspects of the duality simplify: 
\begin{itemize}
    \item The dependence on the 't Hooft coupling $g^2_{\text{YM}}N$ drops out. As a consequence, the computations on the chiral algebra side are essentially free theory computations (when considering BRST closed operators). The only coupling constant is the rank of the gauge group $N$, which gets mapped to the string coupling $g_s\sim 1/N$. Therefore, the large $N$ computations on the gauge theory side can be readily matched with perturbative topological string theory expansion.
    
    \item The closed string field theory of the topological B-model is the Kodaira-Spencer (BCOV) theory \cite{Bershadsky:1993cx,Bershadsky:1993ta}. This theory has been shown to admit a unique quantization in perturbation theory \cite{Costello:2012cy,Costello:2015xsa}.
    
    \item D-branes in the topological B-model are holomorphic submanifolds of the target Calabi-Yau. In particular, D1-branes that will be considered in this note are holomorphic curves in $SL(2,\mathbb{C})$.
    
\end{itemize}

\subsection{Summary of results} 

In \cite{Costello:2018zrm}, the holographic dictionary between single trace operators of the chiral algebra $\mathcal{A}_N$ and modifications of boundary conditions of the Kodaira-Spencer theory on $SL(2,\mathbb{C})$ was proposed. In the following sections, we review two generalization studied in \cite{Budzik:2021fyh,Budzik:2022hcd}:
\begin{enumerate}
    \item The correspondence between determinant operators and ``Giant Graviton" D1-branes.
    \item The duality between the chiral algebra $\mathcal{A}_N$ in non-conformal vacua and topological B-model on asymptotically $SL(2,\mathbb{C})$ geometries.
\end{enumerate}


As part of the standard AdS/CFT correspondence, an insertion of a determinant operator in $\mathcal{N}=4$ SYM is dual to a Giant Graviton D3-brane, which asymptotically wraps an $\mathbb{R}_+ \times S^3$ inside $\textrm{AdS}_5\times S^5$. In twisted holography, determinant operators in the chiral algebra $\mathcal{A}_N$ are dual to D1-branes\footnote{The twisting procedure is expected to reduce D3-branes to D1-branes \cite{Ooguri:1999bv}.} that asymptotically wrap $\mathbb{C}^* \cong \mathbb{R}_+ \times S^1$ inside $SL(2,\mathbb{C})\cong \textrm{AdS}_3\times S^3$. 


When considering insertions of multiple determinants, there might be different brane configurations (eg. connected or disconnected) with the same boundary behavior. Correlation functions of multiple determinants have large $N$ saddles \cite{Jiang:2019xdz}. Using a spectral curve construction, to each large $N$ saddle we can associate a holomorphic curve in $SL(2,\mathbb{C})$, which we conjecture is the support of the dual D1-brane \cite{Budzik:2021fyh}. The conjecture is tested by various holographic computations summarized in the next section.


As the next step in studying twisted holography, in \cite{Budzik:2022hcd} it was proposed that the duality can be extended to non-conformal vacua of the chiral algebra $\mathcal{A}_N$. The holographic dual geometries are deformations of $SL(2,\mathbb{C})$, originating from the backreaction of a stack of non-coincident D1-branes in the topological B-model on $\mathbb{C}^3$.

This is a twisted analog of the duality between the Coulomb branch of $\mathcal{N}=4$ SYM and the (asymptotically $\mathrm{AdS}_5\times S^5$) multi-center supergravity solutions obtained from a near-horizon limit of a stack of non-coincident D3-branes \cite{Douglas:1998tk,Bilal:1998ck,Wu:1998hx,Tseytlin:1998cq,Kraus:1998hv}.

Non-conformal (translation-invariant) vacua of a chiral algebra are a novel observable and in \cite{Budzik:2022hcd} it was conjectured that they correspond to the notion of \emph{the associated variety} of the chiral algebra.\footnote{Chiral algebras are referred to as vertex algebras in the math literature.} In case of chiral algebras of $\mathcal{N}=2$ SCFTs, we expect the 2d vacua to descent from the Higgs branch vacua of the 4d parent theory.\footnote{In a setting without conformal symmetry, the chiral algebra subsector can be defined using a version of the $\Omega$-deformation \cite{Jeong:2019pzg,Butson:2020mmu,Oh:2019bgz}. We expect, but not prove, that this procedure is compatible with the Higgs branch vevs.} The conjecture is therefore in agreement with \cite{Beem:2017ooy} which identified the Higgs branch of an $\mathcal{N}=2$ SCFT with the associated variety of its chiral algebra subsector.

\subsection{Future directions}

Twisted holography setup provides an excellent setting for computations that might be challenging in the full physical duality and for making more precise connections with mathematics.

One natural next step is to study even heavier operators of order $N^2$, such as $(\det Z)^N$, which are expected to produce finite deformations of the geometry \cite{Lin:2004nb}.


In the dual picture, one can consider a stack of $N$ D1-branes wrapping $\mathbb{C}^*$ at $b=c=0$ inside $SL(2,\mathbb{C})$ given by $ad-bc=1$. The backreacted geometry can be described in patches \cite{KBDG}.

The coordinates $b,c$ remain holomorphic and coincide in all patches. In the patch $a\neq 0$, the coordinate $a$ gets deformed to
\begin{equation}
\begin{aligned}
    \tilde a_b  &= a \, e^{-\frac{\bar c}{b} \frac{1}{|b|^2+|c|^2}}, &\quad \text{in } b\neq 0 \text{ patch} \\
    \tilde a_c  &= a \, e^{\frac{\bar b}{c} \frac{1}{|b|^2+|c|^2}}, &\quad \text{in } c\neq 0 \text{ patch}
\end{aligned}
\end{equation}
with coordinate transition
\begin{align}
    \frac{\tilde a_c}{\tilde a_b} = e^{\frac{1}{bc}} \, .
\end{align}
Analogously, in the $d\neq 0$ patch:
\begin{equation}
\begin{aligned}
    \tilde d_b  &= d \, e^{\frac{\bar c}{b} \frac{1}{|b|^2+|c|^2}}, &\quad \text{in } b\neq 0 \text{ patch} \\
    \tilde d_c  &= d \, e^{-\frac{\bar b}{c} \frac{1}{|b|^2+|c|^2}}, &\quad \text{in } c\neq 0 \text{ patch}
\end{aligned}
\end{equation}
with coordinate transition
\begin{align}
    \frac{\tilde d_c}{\tilde d_b} = e^{-\frac{1}{bc}} \, .
\end{align}
And finally, the coordinate transformation between $a\neq 0$ and $d\neq 0$ patches:
\begin{align}
    \tilde a_b \tilde d_b =1 + bc \, .
\end{align}

It would be interesting to perform holographic computations in this setup. One of the challenges is the analysis of determinant correlation functions when the number of determinants is of order $N$.

Finally, we mention some other possible future research directions:
\begin{itemize}
    \item It would be interesting to find the D3-branes in the physical 10d theory that descend to the proposed holomorphic curves in $SL(2,\mathbb{C})$.
    \item It could be conductive to consider determinant operators of adjoint as well as fundamental fields, which holographically translates to including space-filling branes.
    \item Mathematically, the conjecture of \cite{Budzik:2021fyh} can be stated as a correspondence between solutions of matrix equations (\ref{eq:saddle}) and holomorphic curves in $SL(2,\mathbb{C})$. For genus $g=0$, this map was shown to be one-to-one (up to some non-physical redefinitions). It would be interesting to analyze the higher genus cases.
    \item Determinant modifications, reviewed in section \ref{sec:detmod}, form a module for the infinite $N$ global symmetry algebra identified in \cite{Costello:2018zrm}. It could be useful to study this algebra in more detail. For example, a classification of its modules could provide a classification of objects in the topological B-model on $SL(2,\mathbb{C})$.
\end{itemize}


\section{Giant Gravitons and determinants}

In this section, we review the correspondence between determinant operators and Giant Graviton D1-branes in the twisted holography setup which was studied in \cite{Budzik:2021fyh}.


\subsection{Determinant correlation functions} \label{sec:det}

The chiral algebra $\mathcal{A}_N$ is a gauged $\beta\gamma$ system of symplectic bosons $X,Y$ valued in the adjoint representation of $U(N)$. It is convenient to define the linear combination
\begin{align}
    Z(u;z) \equiv X(z) + u Y(z) .
\end{align}
Then the symplectic boson OPE is\footnote{We include the $N^{-1}$ factor in the OPE and also put $N$ in front of single-trace operators. With this choice of conventions, the ribbon diagrams of genus $g$ contribute at order $N^{2-2g}$ in the ’t Hooft expansion.}
\begin{align}
    Z(u;z)^a_b \, Z(v;w)^c_d \sim \delta^a_d \delta_b^c \frac{1}{N} \frac{u-v}{z-w} \, .
\end{align}

We are interested in correlation functions of determinant operators:
\begin{align} \label{eq:det}
    \mathcal{D}(m;u;z) \equiv \det\qty(m+Z(u;z))\, , \quad m\in\mathbb{C} \, .
\end{align}

An insertion of a determinant of this form corresponds to a Giant Graviton brane wrapping a 1-dimensional complex curve, which approaches the boundary of $\text{AdS}_3$ at a point $z$ along the line $b=au-m+\mathcal{O}(a^{-1})$, where $ad-bc=1$ are the coordinates of $SL(2,\mathbb{C}$).\footnote{The boundary behavior of a D1-brane dual to a determinant can be derived by adding a probe D1'-brane transverse to the stack of $N$ D1-branes and finding its image in the backreacted geometry (see \cite{Budzik:2021fyh}).} When considering insertions of multiple determinants $\mathcal{D}(m_i;u_i;z_i)$ there might be many brane configurations with the same asymptotics controlled by parameters $m_i,u_i,z_i$. 



A method of computing correlations functions of determinant operators in the large $N$ limit was presented in \cite{Jiang:2019xdz}. Following their prescription\footnote{Notice that the tree level computations of \cite{Jiang:2019xdz} give an exact answer in the chiral algebra (when considering BRST closed operators) since the 't Hooft coupling dependence drops out in this subsector.} (also implemented in \cite{Berenstein:2022srd, Chen:2019kgc, Yang:2021kot}), we fermionize the determinants:
\begin{align} \label{eq:fermions}
    \expval{\prod_i^k \mathcal{D}(m_i;u_i;z_i)} = \int [\dd\psi\dd\bar\psi] \, \expval{ \prod_i^k e^{\bar\psi_i(m_i+Z(u_i;z_i))\psi^i} } \, ,
\end{align}
where $\psi^i,\bar\psi_i$ are auxiliary (anti)fundamental fermions and $\dd\psi\dd\bar\psi\equiv\prod_i^k \dd \psi^i \dd\bar\psi_i$. The expectation value on the right hand side stands for the chiral algebra path integral. Since the action for the symplectic bosons is free: $N\int\Tr X\bar\partial Y$, we can easily integrate them out:
\begin{align}
     \int [\dd\psi\dd\bar\psi] \, e^{-\frac{1}{2N} \sum_{i\neq j} \frac{u_i-u_j}{z_i-z_j} \bar\psi_i\psi^j\bar\psi_j\psi^i +\sum_i m_i \bar\psi_i\psi^i } \, .
\end{align}
To deal with the above integral, we perform the Hubbard-Stratonovich transformation by introducing auxiliary bosonic variables $\rho^i_j$ for $i\neq j$ (and set $\rho^i_i\equiv m_i$):
\begin{align} \label{eq:rho}
    \frac{1}{Z_\rho}\int [\dd\psi\dd\bar\psi][\dd\rho] \, e^{\frac{N}{2} \sum_{i\neq j} \frac{z_i-z_j}{u_i-u_j} \rho^i_j\rho^j_i +\sum_{i,j} \rho^i_j \bar\psi_i\psi^i } \, ,
\end{align}
where
\begin{align}
    Z_\rho = \int [\dd\rho] \, e^{\frac{N}{2} \sum_{i\neq j} \frac{z_i-z_j}{u_i-u_j} \rho^i_j\rho^j_i}
\end{align}
and the integral $\int[\dd\rho]$ involves only the off-diagonal components of $\rho$ (and an appropriate contour).

After integrating out the fermions, the $\rho$ integral takes a form suitable for a large $N$ saddle-point approximation:
\begin{align}
    \frac{1}{Z_\rho}\int [\dd\rho] \, e^{N S[\rho] } \, ,
\end{align}
with the action
\begin{align} \label{eq:action}
    S[\rho] = \frac{1}{2} \sum_{i\neq j} \frac{z_i-z_j}{u_i-u_j} \rho^i_j \rho^j_i + \log\det \rho \, .
\end{align}
The saddle point equations are
\begin{align}
  \frac{z_i-z_j}{u_i-u_j} \rho^i_j + (\rho^{-1})^i_j = 0, \quad i\neq j \, .  
\end{align}
They can be rewritten in a simple form as a matrix equation:
\begin{align} \label{eq:saddle}
[\zeta,\rho]+[\mu,\rho^{-1}] = 0 \, ,
\end{align}
where
\begin{itemize}
    \item $\rho$ is a $k\times k$ matrix, whose diagonal elements are fixed to be $\rho^i_i=m_i$ and the off-diagonal components are the variables we are solving for,
    \item $\zeta$ is a diagonal $k\times k$ matrix, whose diagonal elements are the positions $z_i$ of determinants on the boundary of $\mathrm{AdS}_3$,
    \item $\mu$ is a diagonal $k\times k$ matrix, whose diagonal elements $u_i$ control the linear combinations of the symplectic bosons $X,Y$ employed in determinants.
\end{itemize}

\subsection{Spectral curve construction}

In \cite{Budzik:2021fyh}, a spectral curve construction for Giant Graviton D1-branes was proposed. To each saddle $\rho$, we can associate a holomorphic curve in $SL(2,\mathbb{C})$ defined as a spectral curve of a system of commuting matrices:
\begin{equation}
\begin{aligned}
    B(a) &= a\mu -\rho \\
    C(a) &= a\zeta +\rho^{-1} \\
    D(a) &=  a\zeta\mu + \rho^{-1}\mu - \zeta\rho \, .
\end{aligned}
\end{equation}

The spectral curve $\mathcal{S}_\rho$ is defined as a set of points $(a,b,c,d)$, where $a\in\mathbb{C}$ and $b,c,d$ are simultaneous eigenvalues of $B(a), C(a), D(a)$.

The matrices are defined in such a way that:
\begin{itemize}
    \item they commute when $\rho$ satisfies the saddle point equations (\ref{eq:saddle}). Therefore, for each saddle $\rho$, they can be simultaneously diagonalized,
    \item they satisfy
    \begin{align}
        aD(a)-B(a)C(a) =1 \, ,
    \end{align}
    which constraints the simultaneous eigenvalues to lay inside the locus $ad-bc=1$. As a result, the spectral curve is a holomorphic curve inside $SL(2,\mathbb{C})$,
    \item the boundary behavior of the spectral curve matches the expected boundary behavior of a D1-brane dual to $k$ insertions of determinants $\mathcal{D}(m_i;u_i;z_i)$ in the boundary theory. Explicitly, when $a\rightarrow\infty$, the eigenvalues of matrices $B(a)$ and $C(a)$ approach
    \begin{equation} \label{eq:pi}
    \begin{aligned}
        b_i &= au_i - m_i + \dots \\
        c_i &= az_i + p_i + \dots \, , 
    \end{aligned}
    \end{equation}
    where $p_i\equiv[\rho^{-1}]^i_i$. This means that the spectral curve reaches the holographic boundary at $k$ points $z_i$, with an asymptotic behavior controlled by $u_i$ and $m_i$ as expected above.
\end{itemize}

\subsection{Holographic checks}

The spectral curve conjecture was tested by matching several chiral algebra and BCOV computations \cite{Budzik:2021fyh}. Here, we summarize three of them.

\subsubsection{Determinant correlation functions with a single trace}

First, we compare the large $N$ expectation value of a single-trace operator in
the presence of multiple determinants with a holographic Witten diagram computation.

A similar calculation to the one in section \ref{sec:det} yields 
\begin{align}
    \expval{N\Tr Z(u;z)^n \prod_i^k \mathcal{D}(u_i;z_i;m_i)} &= \frac{1}{Z_\rho} \int [\dd\rho] e^{N S[\rho]} \, \qty[-N\Tr_{k\times k} \qty(-\rho^{-1}\frac{\mu-u}{\zeta-z})^n ] \, .
\end{align}

The $\rho$ integral is controlled by the same large $N$ saddles (\ref{eq:saddle}). At a given saddle $\rho=\rho^*$, it is equal to evaluating $e^{N S[\rho^*]}$ times
\begin{align} \label{eq:sTr}
    - N \Tr_{k\times k} \qty( -\rho^{-1} \frac{\mu-u}{\zeta-z} )^n \bigg|_{\rho=\rho^*} \, .
\end{align}

The holographic computation corresponds to a closed string propagating from a boundary insertion $N\Tr Z(u;z)^n$ and interacting with a Giant Graviton brane. The BCOV calculation accords to integrating a bulk-to-boundary propagator sourced by $N\Tr Z(u;z)^n$ along the brane:
\begin{align}
    \int_{\mathcal{S}} \partial^{-1} \alpha_n(u;z) \, .
\end{align}
It was shown in \cite{Budzik:2021fyh} that (\ref{eq:sTr}) can be rewritten in the above form as an integral over the spectral curve $\mathcal{S}=S_{\rho^*}$. 

\subsubsection{Matching actions}

An insertion of a determinant operator in the boundary can be engineered in the bulk by placing a probe D1'-brane transverse to the stack of $N$ D1-branes. The auxiliary fermions $\psi^i,\bar\psi_i$, $i=1,\dots,N$ introduced in (\ref{eq:fermions}) can be interpreted as open strings between the probe brane and the stack \cite{Jiang:2019xdz}. The auxiliary variables $\rho^{-1}$ behave as fermion bilinears.

It would be very interesting to match the $\rho$ action (\ref{eq:action}) with the action of open strings on the dual Giant Graviton D1-brane. In \cite{Budzik:2021fyh}, the derivatives of the actions with respect to the parameters $m_i,u_i,z_i$ were matched. In particular, the derivative of the saddle action (\ref{eq:action}) with respect to $m_i$ equals
\begin{align}
    p_i \equiv \pdv{S}{m_i} = [\rho^{-1}]^i_i \, .
\end{align}

The worldvolume theory of a D1-brane is a $u(1)$-gauged $\beta\gamma$ system,\footnote{Not to be confused with the chiral algebra $\mathcal{A}_N$ which is a $\mathrm{u}(N)$-gauged $\beta\gamma$ system, originally appearing as a worldvolume theory of a stack of $N$ D1-branes.} where the two fields, $\beta$ and $\gamma$, correspond to fluctuations in the two transverse directions, which we can take to be $b$ and $c$. The action of the worldvolume theory of a brane $\mathcal{S}$ parametrized by $a$ is\footnote{The denominator comes from contracting with the $SL(2,\mathbb{C})$ volume form, which can be written as $\Omega=\frac{\dd a\dd b \dd c}{a}$ in the $a\neq 0$ patch.}
\begin{align} \label{eq:S}
    \int_{\mathcal{S}} \beta \bar\partial \gamma \wedge \frac{\dd a }{a} \, .
\end{align}
Close to the brane we can expand the two fields into modes:
\begin{align}
    \beta(a) = \sum_n \beta_n a^n \, , \qquad \gamma(a) = \sum_n \gamma_n a^n \, .
\end{align}
The asymptotic behaviour of a brane supported at the spectral curve $\mathcal{S}=\mathcal{S}_{\rho}$ was found to be (\ref{eq:pi}). Near one of the asymptotic boundaries one can then identify the zero-modes as:
\begin{align}
    \beta_0 = -m_i\, , \qquad \gamma_0 = p_i \, .
\end{align}
These zero-modes are conjugate with respect to the action (\ref{eq:S}).

The above considerations give a match of variables $p_i$ defined as conjugate to $m_i$. The derivatives of the actions with respect to $u_i$ and $z_i$ can be matched in a similar but more cumbersome way.

\subsubsection{Modifications of determinants}
\label{sec:detmod}

Excitations of Giant Graviton D1-branes correspond to determinant modifications \cite{Balasubramanian:2004nb,Balasubramanian:2002sa,Berenstein:2003ah,Berenstein:2005vf} such as\footnote{We use a schematic notation:
\begin{align*}
    \varepsilon\varepsilon(Z_1,\dots,Z_N) \equiv \varepsilon_{i_1 \dots i_N}\varepsilon^{j_1\dots j_N} (Z_1)^{i_1}_{j_1} \dots (Z_N)^{i_N}_{j_N} \, .
\end{align*}}
\begin{align}
    \det X = \frac{1}{N!} \varepsilon\varepsilon (X,X,\dots X,X) \mapsto \frac{1}{N!}\varepsilon\varepsilon( X,X,\dots,X,Y^2 ) \, .
\end{align}


One can create BRST-closed modifications of determinants by acting with the modes of the global symmetry algebra of the chiral algebra $\mathcal{A}_N$ in the infinite $N$ limit. This algebra was identified on both sides of the duality in \cite{Costello:2018zrm}. On the chiral algebra side, the global symmetry algebra is defined by certain modes of the BRST-closed single-trace operators:\footnote{There are also three other types of generators of the global symmetry algebra (see \cite{Costello:2018zrm}) but we focus on this one.}
\begin{align}
   a^{(n)}_{p,q} \sim \oint \dd z \, z^{p+\frac{n}{2}} \oint \dd u \, u^{q+\frac{n}{2}} \, \Tr Z(u;z)^n, \quad \abs{p} \leq \frac{n}{2}-1, \,\, \abs{q}\leq \frac{n}{2} \, .
\end{align}
For example, the lowest modes generate an $\mathrm{su}(2)$ subalgebra:
\begin{align}
    a^{(2)}_{0,-1} \sim \oint \dd z \, \Tr X^2(z), \quad a^{(2)}_{0,0} \sim \oint \dd z \, \Tr XY(z), \quad a^{(2)}_{0,1} \sim \oint \dd z \, \Tr Y^2(z) \, .
\end{align}

The global symmetry algebra, when acting on determinants, produces BRST-closed modifications, eg.:
\begin{align}
    \qty[a^{(2)}_{0,1} , \det X(0) ] \sim \varepsilon\varepsilon(X,\dots,X,Y) \, .
\end{align}

In principle, different modes can produce BRST-equivalent operators. The BRST-inequivalent modifications can be identified by studying the following matrix of two-point functions:
\begin{align}
    \expval{ \qty[a^{(n)}_{-p,-q} , \det Y(\infty)] \, \qty[a^{(m)}_{p,q} , \det X(0)] } \Big|_{N\rightarrow\infty} \, .
\end{align}

The two types of inequivalent determinant modifications were found to be:
\begin{align}
    [a^{(n)}_{p,p-1}, \det X(0)] \sim & \,\, n \,\varepsilon\varepsilon(X,\dots,Y^{1-2p})  && , \, p\leq \frac{1}{2} \label{eq:1mod} \\ 
    [a^{(n)}_{p,p+1}, \det X(0)] \sim & \,\, n \, \varepsilon\varepsilon(X,\dots,Y^{-1-2p}\partial X) \label{eq:2mod} \\ 
    &+ n \,\varepsilon\varepsilon(X,\dots,\partial^2 Y^{-3-2p}) && , \, p<-\frac{1}{2} \, . \nonumber
\end{align}

On the holographic side, (the bosonic part of) the global symmetry algebra acts by holomorphic divergence-free vector fields on $SL(2,\mathbb{C})$. The modes of dual global symmetry algebra create excitations of the dual Giant Graviton brane.

The brane dual to the insertions of $\det X(0)$ and $\det Y(\infty)$ is
\begin{align}
    g = \mqty( a & 0 \\ 0 & \frac{1}{a} ) \subset SL(2,\mathbb{C}) \, .
\end{align}
Likewise, we find two types of brane excitations
\begin{align}
    \delta g &= \mqty( a & \delta b \\ 0 & \frac{1}{a}), \quad \delta b \sim n a^{1-2p} \\
    \delta g &= \mqty( a & 0 \\ \delta c & \frac{1}{a}), \quad \delta c \sim n a^{-1-2p} \, ,
\end{align}
which can be matched with the two types of determinant modifications \eqref{eq:1mod} and \eqref{eq:2mod}.

\section{Non-conformal vacua}

In this section, we review the twisted duality in the presence of non-zero vevs of the chiral algebra correlation functions.

\subsection{Coulomb branch geometries}

The translation-invariant vacua of the chiral algebra $\mathcal{A}_N$ can be  parametrized by eigenvalues $(x_i,y_i)$ of $X$ and $Y$ of multiplicity $N_i=\alpha_iN$, $i=1,\dots,n$.

The dual geometries arise from separating the stack of $N$ D1-branes in $\mathbb{C}^3$. The eigenvalues $x_i$ and $y_i$ are identified with the positions $x=x_i$, $y=y_i$ of $N_i$ branes, where $(x,y,z)$ are the coordinates in $\mathbb{C}^3$.


The backreaction of the stack of non-coincident branes was computed in \cite{Budzik:2022hcd}. The resulting geometries can be described using $2^n$ patches of $\mathbb{C}^3$, where for each $i=1,\dots,n$ either $x-x_i$ or $y-y_i$ is non-zero. The coordinates $x$ and $y$ remain holomorphic and coincide in all patches. The coordinate $z$ gets deformed to a new holomorphic coordinate $z_I$ in a patch $I$. The holomorphic coordinate transformation on the intersection of two patches $I$ and $I'$ which differ by the $i$-th choice only ($y\neq y_i$ in $I$ and $x\neq x_i$ in $I'$) is
\begin{align}
    z_I = z_{I'} + \frac{\alpha_i}{(x-x_i)(y-y_i)} \, .
\end{align}

Let us consider two ``extremal" patches: $I_0$ where $y-y_i$ are all non-zero and $I_\infty$ where $x-x_i$ are all non-zero. The coordinate transformation on the intersection takes the form
\begin{align}
    z_0-z_\infty = \sum_{i=1}^n \frac{\alpha_i}{(x-x_i)(y-y_i)} = \sum_{k\geq 0} \sum_{l\geq 0} \frac{1}{x^{k+1}y^{l+1}} \sum_{i=1}^n \alpha_i x_i^k y_i^l \, .
\end{align}
The first term of the sum ($k=l=0$) describes the standard $SL(2,\mathbb{C})$ geometry obtained by a backreaction of coincident branes:\footnote{In coordinates $w_0=z_0x$, $w_\infty=z_\infty y$, we get the familiar $SL(2,\mathbb{C})$ relation: $w_0y-w_\infty x = 1$.}
\begin{align}
    z_0-z_\infty = \frac{1}{xy} \, .
\end{align}
The subsequent terms $\frac{1}{x^{k+1}y^{l+1}}$ are deformations of the $SL(2,\mathbb{C})$ geometry. The coefficients can be identified with the vevs of single-trace operators in the vacuum parametrized by $(x_i,y_i)$:
\begin{align}
    \expval{\, \mathrm{Tr} X^k Y^l \, } = \sum_{i=1}^n N_i x_i^k y_i^l \, .
\end{align}

\subsection{Higgs branch conjecture}



The Coulomb branch of $\mathcal{N}=4$ SYM is simply $\mathbb{R}^{6N}/S_N$, parametrized by the commuting vevs of the six adjoint scalars $\vec\Phi$. In an $\mathcal{N}=2$ language,  it decomposes into the $\mathcal{N}=2$ Coulomb branch and the $\mathcal{N}=2$ Higgs branch, $\mathbb{C}^N/S_N$ and $\mathbb{C}^{2N}/S_N$, respectively. We expect the translation-invariant vacua of the chiral algebra $\mathcal{A}_N$ to descend from the Higgs branch vacua.

More generally, the Higgs branch of $\mathcal{N}=2$ SCFTs has been conjectured to correspond to the notion of the associated variety of its chiral algebra subsector \cite{Beem:2017ooy}.\footnote{See also \cite{Song:2017oew} and the reviews \cite{Arakawa:2017aon,Arakawa:2017fdq}.}

Motivated by the above considerations, in \cite{Budzik:2022hcd} it was conjectured that the space of translation-invariant vacua of a 2d chiral algebra corresponds to the associated variety of the chiral algebra (see \cite{arakawa2012remark} for the definition). The conjecture can be verified in the case of gauged $\beta\gamma$ systems, which arise from Lagrangian $\mathcal{N}=2$ SCFTs.





\bibliographystyle{amsplain}

\bibliography{mono.bib}

\end{document}